\documentclass[journal]{IEEEtran}
\usepackage{booktabs, multirow, tabularx}
\usepackage{graphicx, subfig}
\usepackage{amsmath, amsthm, amsfonts, amssymb, pifont}
\usepackage{cite}

\begin{document}

\title{Collaborative Spectrum Sensing in Cognitive and Intelligent Wireless Networks: An Artificial Intelligence Perspective}

\author{\IEEEauthorblockN{Peng Yi, and Ying-Chang Liang,~\IEEEmembership{Fellow, IEEE}} \\ 
\thanks{Peng Yi is with the National Key Laboratory of Wireless Communications, and the Center for Intelligent Networking and Communications (CINC), University of Electronic Science and Technology of China, Chengdu 611731, China (email: yipengcd@outlook.com).
\par Ying-Chang Liang is with the Institute of Fundamental and Frontier Sciences and the Center for Intelligent Networking and Communications (CINC), University of Electronic Science and Technology of China (UESTC), Chengdu 611731, China (email: liangyc@ieee.org). The corresponding author is Y.-C. Liang.}
}

\maketitle

\begin{abstract}
Artificial intelligence (AI) has become a key enabler for next-generation wireless communication systems, offering powerful tools to cope with the increasing complexity, dynamics, and heterogeneity of modern wireless environments. To illustrate the role and impact of AI in wireless communications, this paper takes collaborative spectrum sensing (CSS) in cognitive and intelligent wireless networks as a representative application and surveys recent advances from an AI perspective. We first introduce the fundamentals of CSS, including the general framework, classical detector design, fusion strategies and evaluation metrics. Then, we present an overview of the state-of-the-art research on AI-driven CSS, classified into three categories according to learning paradigms: discriminative deep learning (DL), generative DL models, and deep reinforcement learning (DRL). Building on this, we explore AI-empowered semantic communication (SemCom) as a paradigm-shifting solution for CSS. By extracting and transmitting task-relevant features, SemCom upgrades CSS from a computation-centric approach to a highly efficient joint communication and computation framework. Both single-user and multi-user SemCom scenarios are elaborated in detail. Finally, we discuss limitations, open challenges, and future research directions at the intersection of AI and wireless communication.
\end{abstract}


\section{Introduction}
\label{Sec.Introduction}

With the rapid expansion of diverse wireless applications, communication technologies are undergoing revolutionary transformations. Unlike previous generations that primarily focused on enhancing user data rates, the forthcoming sixth-generation (6G) mobile communication system is expected to support highly dynamic, large-scale, and heterogeneous networks characterized by massive device connectivity, ubiquitous coverage, and stringent requirements on latency, reliability, and energy efficiency~\cite{wc/Chen20,comsur/Wang23}. This unprecedented combination of system complexity, performance demands, and wireless channel uncertainty renders conventional design and optimization approaches increasingly inadequate~\cite{comsur/Nguyen26}. 
In parallel, artificial intelligence (AI), particularly deep learning (DL), has demonstrated transformative success in data-intensive fields, most notably in computer vision (CV) and natural language processing (NLP)~\cite{nature/LeCun15}. These achievements establish AI as a general-purpose paradigm for learning intrinsic data characteristics, enabling pattern recognition, content generation, and sequential decision-making. 
Motivated by these advances, the wireless research community has increasingly integrated AI into the design of future wireless communication systems, driving the evolution from the Internet of Things (IoT) toward connected intelligence~\cite{cm/Letaie19}. AI already serves as a fundamental enabler for a wide range of critical 6G functions, including signal detection~\cite{tvt/LiuZ22}, multiple access~\cite{pieee/Cao24}, and resource allocation~\cite{twc/Cao24}, among others.

\begin{table*}[t]
\caption{Comparison of Contributions Between Existing Surveys and This Paper}
\label{tab:comparison}
\centering
\renewcommand{\arraystretch}{1.3}
\small 
    \begin{tabular}{c c c c c c p{8.2cm}} 
    \toprule
    \multirow{2}{*}{\textbf{Ref.}} & \multirow{2}{*}{\textbf{Year}} & \multicolumn{3}{c}{\textbf{AI Methodologies}} & \multirow{2}{*}{\textbf{SemCom}} & \multirow{2}{*}{\textbf{Contributions}} \\
    \cmidrule(lr){3-5} 
    & & \textbf{Dis. DL} & \textbf{Gen. DL} & \textbf{DRL} & & \\
    \midrule \relax
    \cite{wispnet/Muthukumar22} & 2022 & $\checkmark$ & \ding{55} & \ding{55} & \ding{55}  & Compares the performance of prominent DL models evaluated under generalized fading channels.\\
    \cite{access/Syed23} & 2023 & $\checkmark$ & \ding{55} & \ding{55} & \ding{55}  & Reviews discriminative DL architectures for CSS and summarizes relevant radio frequency signal datasets.\\
    \cite{icoin/Pham23} & 2023 & \ding{55} & \ding{55} & $\checkmark$ & \ding{55}  & Investigates DRL-based sensing policies to balance energy-performance trade-offs in CR networks.\\
    \cite{pieee/Cao24} & 2024 & $\checkmark$ & \ding{55} & $\checkmark$ & \ding{55} & Highlights the fundamental role of AI-driven CSS in multiple access.\\
    \cite{cn/Falco26} & 2026 & $\checkmark$ & \ding{55} & $\checkmark$ & \ding{55} & Synthesizes machine learning and DL trends in CSS through a meta-analysis of over $100$ existing studies.\\
    \midrule \relax
    \textbf{This paper} &  & $\checkmark$ & $\checkmark$ & $\checkmark$ & $\checkmark$ & \textbf{Provides a comprehensive review of AI-driven CSS across three learning paradigms and presents the first survey of SemCom-enabled CSS from a joint communication-computation perspective.}\\
    \bottomrule
    \end{tabular}
    \par\vspace{4pt}
    \begin{minipage}{\textwidth}
        \footnotesize\raggedright
        \emph{Abbreviations:} Ref.: reference, Dis. DL: discriminative deep learning, Gen. DL: generative deep learning, DRL: deep reinforcement learning, SemCom: semantic communication. 
    \end{minipage}
\end{table*}

\begin{figure}[!t]
    \centering
    \includegraphics[width=0.9\linewidth]{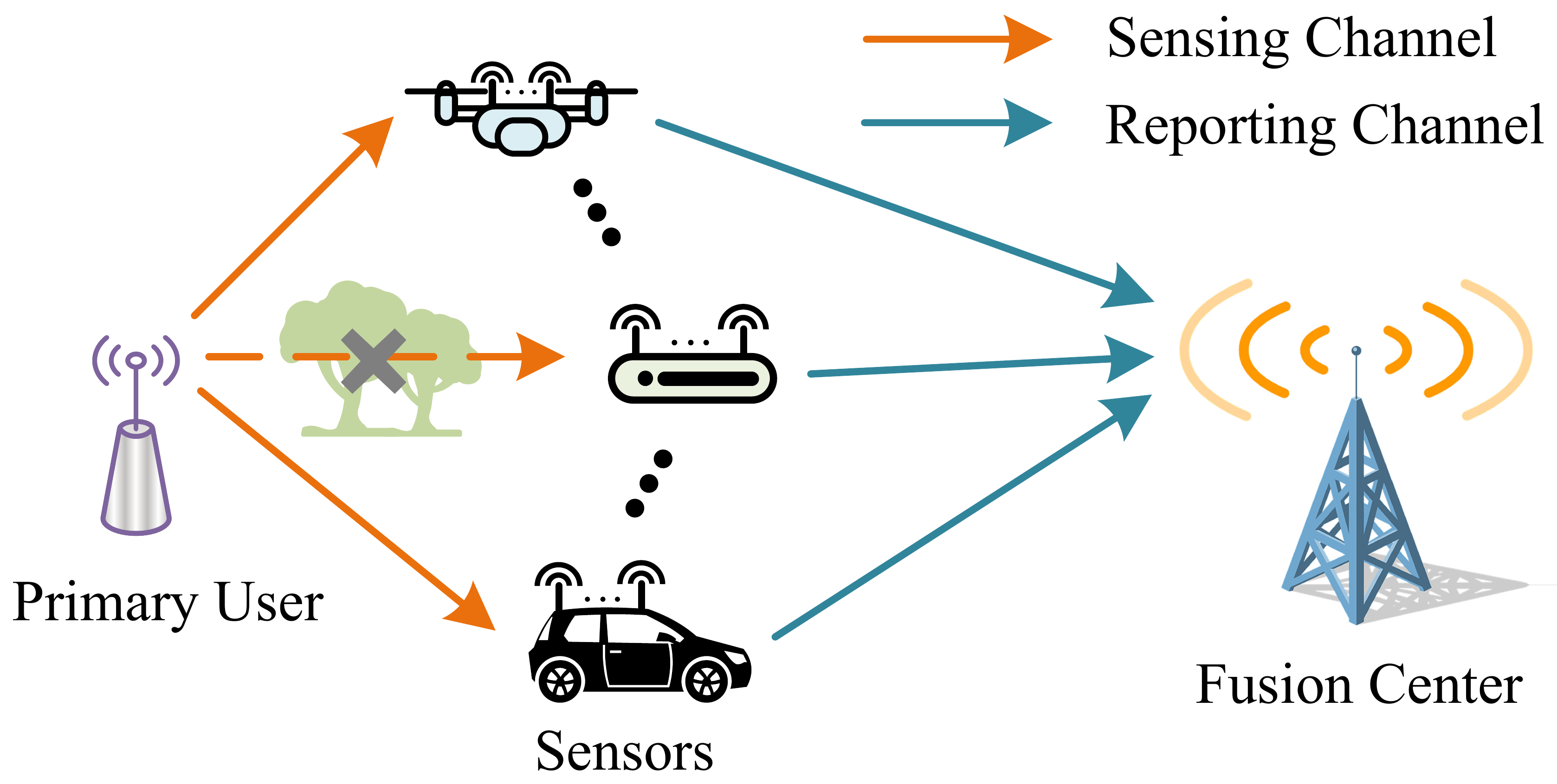}
    \caption{Illustration of the collaborative spectrum sensing scenario.}
    \label{fig:scenario}
\end{figure}

Among these AI-enabled functionalities, collaborative spectrum sensing (CSS) in cognitive and intelligent wireless networks stands out as a representative and practically important use case, as it inherently involves distributed perception, information fusion, and inference under partial observations~\cite{Letaief09,tvt/Liang11,tccn/Qin20}.
As illustrated in Fig.\,\ref{fig:scenario}, CSS is a distributed spectrum sensing paradigm in which multiple cognitive radio (CR) sensors collaboratively detect the presence or absence of primary users (PUs) in a shared spectrum band. In a typical CSS framework, each sensor independently performs local sensing based on its own observations and subsequently reports its sensing outcomes to a fusion center (FC), which makes a global decision by aggregating information from multiple nodes. By exploiting spatial diversity across geographically distributed sensors, CSS can effectively mitigate the impact of adverse propagation conditions and sensing uncertainties, including channel fading, shadowing, and the hidden terminal problem, thereby improving detection reliability compared to individual sensing~\cite{Letaief09,comsur/Cichon16}. Such reliable spectrum awareness provides the foundation for opportunistic spectrum access (OSA), enabling secondary users (SUs) to efficiently exploit underutilized spectrum resources without causing harmful interference to PUs~\cite{wc/MitolaM99}. However, the performance of CSS critically depends on how local observations are represented, communicated, and fused under practical constraints, such as model mismatch arising from channel uncertainty, limited reporting bandwidth, and latency~\cite{comsur/AliH17}.

These challenges have motivated growing interest in applying AI techniques to CSS, as learning-based approaches offer new opportunities to cope with sensing uncertainty, communication constraints, and large-scale collaboration. In recent years, CSS has increasingly been revisited from an AI perspective, marking a paradigm shift from model-based designs toward data-driven approaches~\cite{access/Syed23}. Conventional CSS methods typically rely on explicit signal and channel models, handcrafted features, and analytically derived thresholds and fusion rules, which restrict their flexibility in complex and dynamic wireless environments~\cite{jsac/Liu19}. By contrast, AI-enabled CSS leverages DL to automatically extract informative sensing features, adapt fusion strategies, and support scalable collaboration among distributed sensors. 
Specifically, existing AI-driven CSS methodologies can be systematically categorized into three major learning paradigms: 1) \textit{discriminative DL}, which excels at extracting high-dimensional sensing features and establishing complex decision boundaries for precise PU state classification; 2) \textit{generative DL}, which is capable of modeling underlying data distributions to augment scarce sensing samples and mitigate adversarial channel effects; and 3) \textit{deep reinforcement learning (DRL)}, which enables cognitive nodes to dynamically learn and optimize sequential sensing and reporting policies through continuous interaction with the uncertain radio environment.

Furthermore, despite the remarkable success of AI-driven CSS methodologies, most existing works treat CSS merely as an isolated computation (i.e., pattern recognition) task, thereby overlooking the fundamental interplay between sensing data processing/fusion and limited reporting channel bandwidth. Building upon existing AI techniques, the emerging paradigm of semantic communication (SemCom) empowers cognitive nodes to extract and exchange only the task-critical features that are semantically relevant to the spectrum occupancy detection, rather than transmitting raw or simply compressed data. This paradigm shift intrinsically upgrades CSS from a computation-centric approach to a joint communication and computation framework, offering improved detection accuracy and unprecedented robustness against severe reporting channel impairments. Consequently, SemCom-enabled CSS approaches stand as the next evolutionary step beyond conventional AI-driven CSS approaches, establishing a novel framework where modern AI algorithms and semantic-aware communication architectures jointly transform wireless sensing and collaborative inference.

\subsection{Comparisons and Contributions}
Prior to our work, several existing surveys have summarized the latest developments in AI-driven CSS methodologies. Ref.\,\cite{access/Syed23} provide a comprehensive review of conventional machine learning and early discriminative DL models for CSS, while also introducing relevant radio frequency signal datasets. Treating SUs as intelligent agents, Ref.\,\cite{icoin/Pham23} focus on the application of DRL approaches to construct sensing policies that balance performance-energy trade-offs in CR networks. Ref.\,\cite{pieee/Cao24} investigate CSS from the perspective of multiple access, outlining relevant discriminative DL and DRL schemes as CSS is the fundamental prerequisite for dynamic channel access and subsequent data transmission. More recently, Ref.\,\cite{cn/Falco26} conduct a meta-analysis of over $100$ studies to synthesize the applications of machine learning, discriminative DL, and DRL in CSS. Their study also identifies overarching trends and highlights promising future directions.

Despite these valuable contributions, previous surveys only partially cover the spectrum of AI learning paradigms, and the potential of generative DL has been largely overlooked. To bridge this gap, we aim to provide a comprehensive overview of AI-driven CSS across diverse learning paradigms, explicitly incorporating generative DL alongside discriminative DL and DRL. Moreover, the emerging paradigm of SemCom in CSS remains entirely unexplored in prior studies. Therefore, we introduce a novel perspective by examining SemCom-enabled CSS, which reconceptualizes CSS from an isolated computation task into a joint communication and computation framework. Building upon this, we categorize SemCom-enabled CSS into single-user and multi-user scenarios and investigate each respectively. Tab.\,\ref{tab:comparison} presents a detailed comparison between our work and existing surveys.

\subsection{Contents and Structures}
\begin{figure}
    \centering
    \includegraphics[width=0.95\linewidth]{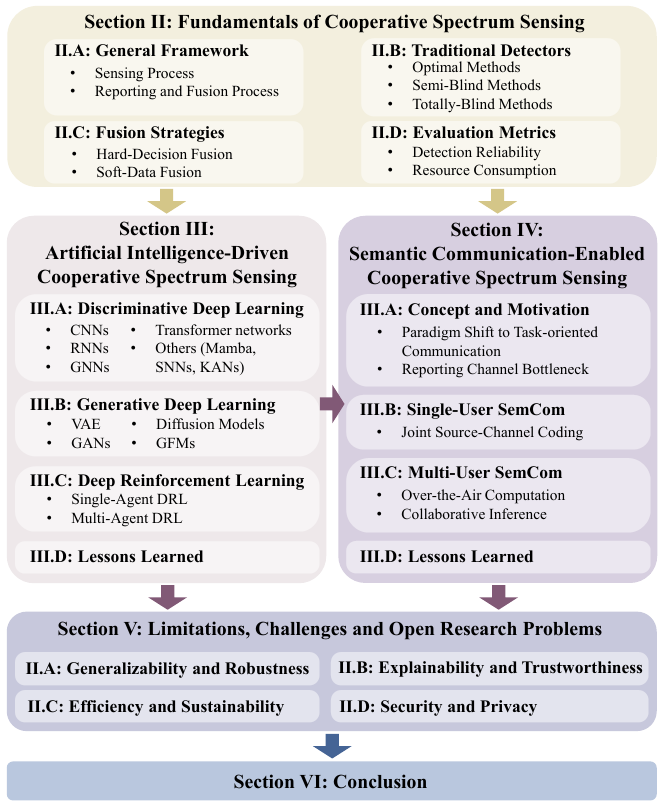}
    \caption{Organization and structure of this paper.}
    \label{fig:structure}
\end{figure}

The organization of this paper, as illustrated in Fig.\,\ref{fig:structure}, is as follows. Section\,\ref{Sec.Foundation} outlines the fundamentals of CSS, introducing the general framework, traditional detectors, fusion strategies, and evaluation metrics. Subsequently, Section\,\ref{Sec.AI} provides a comprehensive taxonomy of AI-driven CSS methodologies, categorizing existing approaches into discriminative DL, generative DL, and DRL. From the perspective of joint communication and computation, Section\,\ref{Sec.SemCom} introduces and explores SemCom-enabled CSS, covering both single-user and multi-user scenarios. Finally, Section\,\ref{Sec.Challenges} discusses future challenges and open research directions, and Section\,\ref{Sec.Conclusion} concludes the paper.

\section{Fundamentals of Cooperative Spectrum Sensing}
\label{Sec.Foundation}

\begin{figure*}[t]
    \centering
    \includegraphics[width=0.9\linewidth]{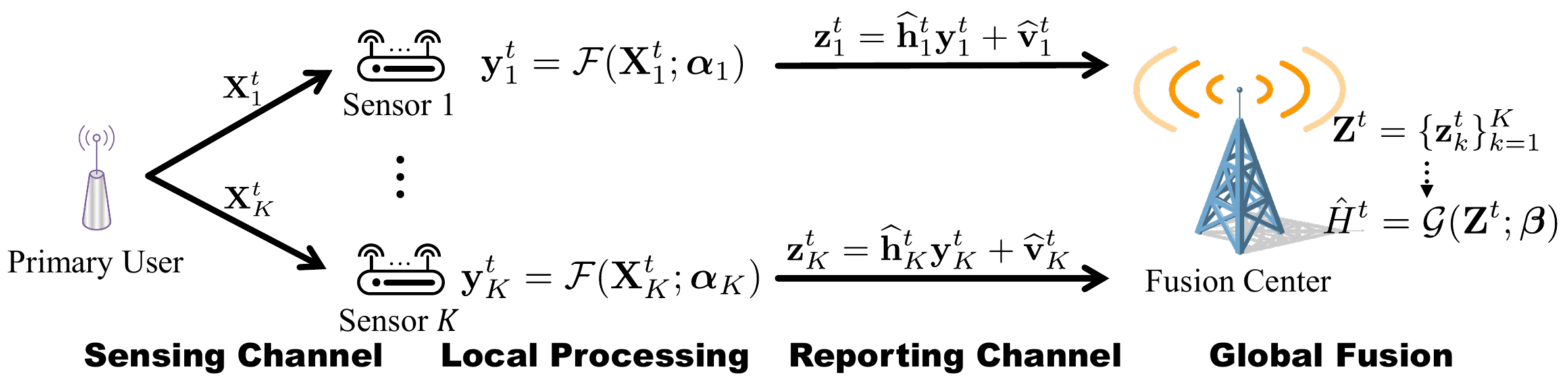}
    \caption{System-level diagram of the collaborative spectrum sensing framework, illustrating the end-to-end signal processing and information fusion workflow.}
    \label{fig:System}
\end{figure*}

\subsection{General Framework}
\label{SubSec.GeneralFramework}
\begin{figure}[t]
    \centering
    \includegraphics[width=0.8\linewidth]{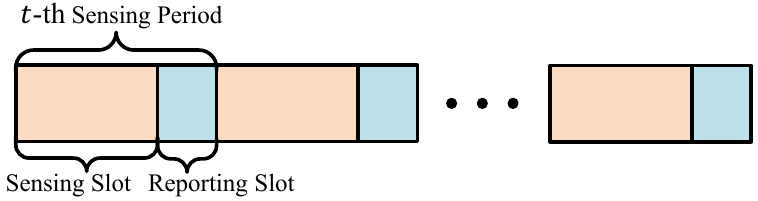}
    \caption{Timing structure of the collaborative spectrum sensing process. Each sensing period consists of a sensing slot followed by a reporting slot.}
    \label{fig:timeslot}
\end{figure}

To summarize the overall architecture, Fig.\,\ref{fig:System} presents a unified CSS framework designed to be compatible with both conventional model-based designs and AI-driven approaches. As depicted, this versatile structure captures the complete signal processing workflow, regardless of whether the underlying functions are implemented via mathematical modeling or neural networks.
Specifically, $K$ spatially distributed sensors collaboratively monitor the spectrum to infer the occupancy status of a PU. In practice, CSS operates over successive sensing periods indexed by $t$. As shown in Fig.\,\ref{fig:timeslot}, each sensing period consists of one sensing slot followed by one reporting slot. Within the $t$-th sensing slot, each sensor equipped with $M$ antennas collects $N$ baseband samples, indexed by $n\in\{1,\ldots,N\}$. For the $k$-th sensor, the received baseband sample $\mathbf{x}_k^{t}(n) \in \mathbb{C}^{M \times 1}$ can be formulated as a binary hypothesis testing problem:
\begin{equation}
    \label{eq:hypothesis}
    \mathbf{x}_k^{t}(n) =
    \begin{cases}
        \widetilde{\mathbf{h}}_k^{t} s^{t}(n) + \widetilde{\mathbf{u}}_k^{t}(n), & H_1, \\
        \widetilde{\mathbf{u}}_k^{t}(n), & H_0,
    \end{cases}
\end{equation}
where $s^{t}(n)$ denotes the PU signal, $\widetilde{\mathbf{h}}_k^{t} \in \mathbb{C}^{M \times 1}$ represents the sensing channel gain vector between the PU and the $k$-th sensor, which is assumed to be constant during each sensing period, and $\widetilde{\mathbf{u}}_k^{t}(n) \in \mathbb{C}^{M \times 1}$ is the additive noise vector. The hypotheses $H_1$ and $H_0$ correspond to the presence and absence of the PU, respectively.

Let $\mathcal{F}(\cdot;\boldsymbol{\alpha}_k)$ denote the local sensing and processing function at the $k$-th sensor, parameterized by $\boldsymbol{\alpha}_k$. Here, $\boldsymbol{\alpha}_k$ encapsulates the algorithm parameters of local processing, which may correspond to detector settings in conventional methods or learned model parameters in data-driven approaches.
Based on its local observation sequence $\mathbf{X}_k^{t} = {\{\mathbf{x}_k^{t}(n)\}}_{n=1}^N$, each sensor generates a message
\begin{equation}
    \label{eq:encoding}
    \mathbf{y}_k^{t} = \mathcal{F}(\mathbf{X}_k^{t}; \boldsymbol{\alpha}_k),
\end{equation}
where $\mathbf{y}_k^{t}$ represents the information to be reported in the reporting slot $t$. Depending on the specific design, $\mathbf{y}_k^{t}$ may correspond to a binary decision, a sufficient statistic, or a task-oriented semantic representation.

This message is conveyed to the FC through the reporting link, denoted as $\mathcal{C}(\cdot)$, which is subject to channel impairments. A generic reporting model can be expressed as
\begin{equation}
    \label{eq:report}
    \mathbf{z}_k^{t} = \mathcal{C}(\mathbf{y}_k^{t}) = \widehat{\mathbf{h}}_k^{t} \mathbf{y}_k^{t} + \widehat{\mathbf{v}}_k^{t},
\end{equation}
where $\widehat{\mathbf{h}}_k^{t}$ and $\widehat{\mathbf{v}}_k^{t}$ denote the reporting channel gain and noise in the reporting slot $t$, respectively.

By denoting the aggregated received messages from all sensors at the FC in reporting slot $t$ as $\mathbf{Z}^{t} = \{\mathbf{z}_k^{t}\}_{k=1}^K$, a global decision on the spectrum occupancy is obtained through a fusion function
\begin{equation}
    \label{eq:decoding}
    \hat{H}^{t} = \mathcal{G}(\mathbf{Z}^{t}; \boldsymbol{\beta}),
\end{equation}
where $\mathcal{G}(\cdot;\boldsymbol{\beta})$ represents the fusion or decoding strategy with parameters $\boldsymbol{\beta}$, encompassing conventional fusion rules, learning-based fusion networks, as well as semantic decoders. Over successive sensing periods, the system can adapt its sensing operations (e.g., sensor participation, sensing and reporting resource allocation, and sensing channel selection) based on historical sensing outcomes (e.g., $\hat{H}^{t}$) and operational objectives. As a result, decisions and feedback in period $t$ influence the configuration and observations in period $t{+}1$, naturally leading to sequential decision-making formulations.

\subsection{Traditional Detectors}
Over the past decades, a variety of model-based spectrum sensing detectors have been developed to enable OSA in cognitive and intelligent wireless networks~\cite{comsur/AliH17}. Within the general CSS framework introduced in Section\,\ref{SubSec.GeneralFramework}, these classical detectors can be viewed as specific realizations of the local processing function $\mathcal{F}(\cdot;\boldsymbol{\alpha})$, where $\boldsymbol{\alpha}$ is determined by analytically derived detection rules. Depending on their reliance on prior knowledge of the PU signal and noise statistics, traditional detectors are commonly categorized into optimal, semi-blind, and totally-blind methods~\cite{jsac/Liu19}.

\subsubsection{Optimal Methods}
Optimal detectors are derived based on the likelihood ratio test (LRT), which maximizes the detection probability for a given false alarm constraint. A representative example is the estimator-correlator (E-C) detector, which achieves the theoretical performance upper bound when the statistical covariance matrices of both the PU signal and the noise are perfectly known. In the context of CSS, such detectors offer valuable performance benchmarks. However, their optimality critically depends on comprehensive prior knowledge, including signal structure and noise statistics, which is rarely available in practical scenarios due to the lack of cooperation between PUs and SUs, as well as the time-varying nature of wireless environments. Consequently, the applicability of optimal detectors in real-world CSS systems is often limited.

\subsubsection{Semi-Blind Methods}
To circumvent the stringent prior knowledge requirements of optimal detectors, semi-blind methods exploit partial statistical information about either the noise or the PU signal. Representative examples include energy detection (ED)~\cite{tcom/DighamAS07} and maximum-eigenvalue detection (MED)~\cite{icc/ZengKL08}, which require knowledge of the noise power, as well as matched filtering (MF)~\cite{globalsip/ZhangCG14} and feature detection (FD)~\cite{jsac/AxellL11}, which exploit prior information about the PU signal or its distinctive features. 
Specifically, ED operates by comparing the received signal energy against a predefined threshold and is particularly attractive due to its low computational complexity and ease of implementation. MED leverages the correlation structure of the received signal covariance matrix and is more effective in detecting correlated PU signals, making it well suited for multi-antenna and cooperative sensing scenarios. MF and FD detect the presence of the PU by correlating the received signal with a known reference waveform or a feature vector, respectively, and can typically achieve relatively high detection performance when accurate PU information is available. 
Despite their widespread adoption, the performance of semi-blind detectors is highly sensitive to the accuracy of noise power estimation or the availability of reliable PU prior knowledge. In the presence of noise uncertainty, imperfect signal models, or time-varying environments, estimation errors and model mismatch can lead to severe performance degradation. These limitations fundamentally constrain the robustness of semi-blind methods in dynamic and heterogeneous sensing environments, motivating the exploration of totally-blind methods.

\subsubsection{Totally-Blind Methods}
Totally-blind detectors are designed to further enhance robustness by eliminating the need for explicit prior knowledge of both PU signals and noise statistics. Representative approaches include maximum-minimum eigenvalue detection (MMED)~\cite{tcom/ZengL09} and covariance absolute value (CAV) detection~\cite{tvt/ZengL09}, which leverage intrinsic structural properties of the received signal covariance matrix. By relying on relative or normalized statistics, these methods exhibit strong robustness to noise uncertainty and are well suited for environments with poorly characterized noise conditions. However, this robustness typically comes at the expense of detection performance. When accurate noise knowledge is available, totally-blind detectors generally underperform compared to semi-blind or optimal methods. This trade-off highlights the inherent tension between robustness and detection accuracy in model-based spectrum sensing.

In summary, the above categories illustrate that classical spectrum sensing detectors correspond to fixed and manually designed $\mathcal{F}(\cdot;\boldsymbol{\alpha})$, each incorporating different assumptions and performance trade-offs. While effective under specific conditions, their limited adaptability motivates the exploration of learning-based approaches that can automatically infer sensing representations from data, as discussed in the next section.

\subsection{Fusion Strategies}
Within the general CSS framework introduced in Section\,\ref{SubSec.GeneralFramework}, the fusion strategy $\mathcal{G}(\cdot;\boldsymbol{\beta})$ plays a central role in aggregating the messages reported by distributed sensors to derive a reliable global decision on spectrum occupancy. Conceptually, different fusion strategies correspond to distinct design choices regarding the form and granularity of the reported information $\mathbf{y}_k$. Based on this criterion, traditional fusion schemes are broadly classified into hard-decision fusion (HDF) and soft-data fusion (SDF).

\subsubsection{Hard-Decision Fusion}
In HDF, each sensor independently performs a local binary decision on the presence of the PU and reports only a one-bit outcome to the FC. The FC then applies simple logical rules, such as the ``AND'' rule, ``OR'' rule, or majority voting, to obtain the final decision. Owing to its extremely low communication overhead and minimal processing complexity, HDF is attractive for bandwidth-constrained sensing networks. However, this simplicity comes at the cost of discarding rich local sensing information during the quantization process, which inevitably leads to performance degradation. 

\subsubsection{Soft-Data Fusion}
In contrast, SDF allows sensors to transmit more informative messages, such as raw signal samples, sufficient statistics, or likelihood metrics, to the FC. By preserving fine-grained sensing information, the FC can exploit spatial diversity more effectively and achieve superior detection performance, often approaching that of centralized optimal detectors. Nevertheless, the improved accuracy of SDF is accompanied by substantially increased reporting channel bandwidth requirements and heightened susceptibility to channel impairments, which may compromise its theoretical performance advantages in practical deployments. 

Overall, traditional fusion strategies reveal a fundamental trade-off in CSS: HDF is limited by information loss, whereas SDF is constrained by communication overhead and reporting channel reliability. These approaches can be viewed as fixed and manually designed realizations of the fusion function $\mathcal{G}(\cdot;\boldsymbol{\beta})$, which lack the flexibility to adaptively balance detection performance and communication efficiency. This observation motivates the development of intelligent and learning-based fusion paradigms, in which the representation, transmission, and fusion of sensing information can be jointly optimized in a task-oriented manner.

\subsection{Evaluation Metrics}
Performance evaluation of CSS schemes generally focuses on two key dimensions: detection reliability and resource consumption. In terms of reliability, the receiver operating characteristic (ROC) curve is the standard metric for characterizing the intrinsic trade-off between the probability of detection ($P_d$) and the probability of false alarm ($P_{fa}$). Here, $P_d$ signifies the probability of accurately identifying an active PU, whereas $P_{fa}$ refers to the probability of incorrectly triggering a detection when the spectrum is idle. Furthermore, the area under the ROC curve (AUC) serves as a scalar metric to quantify the overall sensing performance. 
Beyond reliability, assessing resource consumption is also important. This includes the computational complexity associated with detection algorithms and the communication overhead necessitated by reporting sensing information to the FC. Together, these metrics determine the detection effectiveness and resource efficiency of CSS solutions, ensuring their suitability for deployment in resource-limited wireless networks.

Beyond binary detection, the CSS framework can be extended to spectrum segmentation, a task that entails identifying and localizing signal activities within two-dimensional time-frequency grids. To accommodate this increased granularity, the evaluation metrics further incorporate performance metrics inherited from CV, such as pixel-wise accuracy, precision, recall, and intersection over union (IoU). These metrics enable a more effective assessment of spectrum sensing by quantifying the temporal and spatial alignment between the detected regions and the actual spectrum occupancy.

\section{Artificial Intelligence-Driven Cooperative Spectrum Sensing}
\label{Sec.AI}
In this section, we summarize and categorize representative AI models for CSS, with an emphasis on the underlying learning paradigms and their roles in improving sensing reliability, as illustrated in Tab.\,\ref{tab:taxonomy}. Under the AI-driven formulation, CSS is commonly cast as a binary classification task (or hypothesis testing), in which the two classes correspond to the absence of a PU (null hypothesis $H_0$) and the presence of a PU (alternative hypothesis $H_1$).

\begin{table*}[t]
    \centering
    \caption{Taxonomy of AI-driven CSS models and their roles.}
    \label{tab:taxonomy}
    \begingroup
    \belowrulesep=0pt
    \aboverulesep=0pt
    \doublerulesep 2.2pt
    \footnotesize
    \tabcolsep 4.2pt
    \renewcommand{\arraystretch}{1.28}
    \renewcommand{\tabularxcolumn}[1]{>{\arraybackslash}m{#1}}
    \begin{tabularx}{\textwidth}
    {@{\hspace*{3pt}}@{\extracolsep{\fill}}
    >{\hsize=0.55\hsize\raggedright\arraybackslash}X
    >{\hsize=1.05\hsize\raggedright\arraybackslash}X
    >{\hsize=1.40\hsize\raggedright\arraybackslash}X
    @{\hspace*{3pt}}}
        \toprule
        \textbf{Category} & \textbf{Representative models} & \textbf{Typical roles in CSS} \\
        \hline
        Discriminative DL & CNNs~\cite{jsac/Liu19,zheng2020spectrum,wcl/Cai22}, RNN/LSTM~\cite{vtc/Xu20,access/Soni20}, GNNs~\cite{tvt/Janu23,tccn/Dong25}, Transformer networks~\cite{icl/Chen25,wcl/Fang25}, Mamba~\cite{mlsp/Yi24}, SNNs~\cite{tgcn/Liu24,taes/Dakic24}, KAN~\cite{Pan_2024,comsnets/Patel25}. 
        & Local feature extraction~\cite{jsac/Liu19,jcin/Liu21,wcl/Cai22}; temporal dependency modeling~\cite{vtc/Xu20,mlsp/Yi24}; topology-aware fusion~\cite{tvt/Janu23,tccn/Dong25}; cooperative feature aggregation~\cite{tvt/LeeKC19,wcl/Fang25}; spectrogram segmentation~\cite{wcl/Nguyen25, wcl/Le25, wcl/HuynhThe25}, energy-efficient sensing~\cite{tgcn/Liu24,Pan_2024}. \\
        \hline
        Generative DL & VAE~\cite{tvt/Xie20,tccn/Liao24}, GAN~\cite{icc/Davaslioglu18,jsen/Yan25}, diffusion models~\cite{wcl/HuynhThe26}, GFMs~\cite{alikhani2025lwm,jsca/Zhou25} 
        & Unsupervised representation learning~\cite{tvt/Xie20,tccn/Liao24}; data augmentation and denoising~\cite{icc/Davaslioglu18,jsen/Yan25}; domain adaptation~\cite{icl/Zhao23}; spectrogram segmentation~\cite{wcl/HuynhThe26}; spectrum data understanding~\cite{alikhani2025lwm,jsca/Zhou25}. \\
        \hline
        DRL & DQN-family~\cite{icl/Sarikhani20,ieicetb/Xu25}, Q-learning~\cite{wcl/Ngo23},  MADDPG~\cite{icl/Gao21,tvt/Gao24}, MAPPO~\cite{vtc/Wang25}. 
        & Sensor selection and cooperation control~\cite{icl/Sarikhani20,ieicetb/Xu25}; sensing channel selection~\cite{icl/Gao21,tvt/Gao24,vtc/Wang25}; multi-slot spectrum occupancy prediction~\cite{wcl/Ngo23}. \\
        \hline
        \bottomrule
    \end{tabularx}
    \par\vspace{2pt}
    \begin{minipage}{\textwidth}
        \footnotesize\raggedright\emph{Abbreviations:} DL: deep learning, DRL: deep reinforcement learning, CNNs: convolutional neural networks, RNNs: recurrent neural networks, LSTM: long short-term memory, GNNs: graph neural networks, SNNs: spiking neural networks, KANs: Kolmogorov--Arnold networks, VAE: variational autoencoder, GAN: generative adversarial network, GFMs: generative foundation models, DQN: deep Q-network, MADDPG: multi-agent deep deterministic policy gradient, MAPPO: multi-agent proximal policy optimization.
    \end{minipage}
    \endgroup
\end{table*}

\subsection{Discriminative Deep Learning}
Discriminative DL models constitute the most extensively investigated class of AI techniques for CSS. Beyond handcrafted features and model-based detectors, they typically learn a decision rule by directly extracting discriminative representations from sensing observations (i.e., approximating the posterior $P(H_i|\mathbf{X})$). By leveraging data-driven nonlinear function approximation, these models can capture complex signal characteristics under non-ideal sensing conditions.

\subsubsection{Convolutional Neural Networks}
Convolutional neural networks (CNNs) are among the earliest and most widely adopted discriminative DL models for CSS. While early studies explored directly feeding raw in-phase and quadrature (I/Q) samples (i.e., $\mathbf{X}_k^t$) into CNNs~\cite{zheng2020spectrum,twc/Mehrabian23,iotj/Zhang24}, a more robust paradigm has emerged that transforms these sequences into structured representations to better expose decision-relevant statistics. The most prominent example is the sample covariance matrix at the $k$-th sensor, given by
\begin{equation}
    \mathbf{R}_k^{t} = \frac{1}{N} \sum_{n=1}^{N} \mathbf{x}_k^{t}(n) {\mathbf{x}_k^{t}}^H (n),
\end{equation}
which encapsulates rich spatial and correlation information~\cite{jsac/Liu19,jcin/Liu21}. By interpreting $\mathbf{R}_k^{t}$ as an image-like input, CNNs can effectively extract local patterns related to energy distribution, antenna correlations, and signal subspace structures~\cite{jsac/Liu19}. Consequently, these approaches exhibit superior robustness against noise uncertainty and channel impairments compared to traditional detectors, particularly in multi-antenna scenarios~\cite{icc/Liu19,icct/Liu22}.
To extend this spatial capability to the temporal domain, several works~\cite{icl/Xie19} stack covariance matrices from consecutive sensing periods as the CNN input. This design enables the network to implicitly learn temporal dependencies of PU activities, thereby further improving detection accuracy. Beyond local sensing, CNNs have also been employed for centralized fusion, in which sensing outcomes from multiple sensors are organized into a global matrix representation for joint inference~\cite{tvt/LeeKC19}.
Expanding the scope from narrowband to wideband spectrum sensing, spectrograms provide a natural two-dimensional representation that captures dynamics across both time and frequency domains~\cite{tccn/Lees19,icl/Chen21,wcl/Cai22}. In this context, CNN-based models can also be trained to infer fine-grained occupancy maps, effectively formulating wideband sensing as a semantic segmentation task~\cite{wcl/Nguyen25,wcl/Le25}. Although this formulation transcends conventional binary hypothesis testing, the resulting segmentation outputs can be readily reduced to band-wise occupancy decisions, remaining fully compatible with standard spectrum sensing objectives.

\subsubsection{Recurrent Neural Networks}
To better exploit temporal correlations in PU activity, recurrent neural networks (RNNs), including long short-term memory (LSTM) networks, have been introduced for spectrum sensing. Unlike CNN-based methods that primarily focus on spatial structures, RNNs and LSTMs operate directly on sequential sensing observations, enabling the learning of temporal dependencies across sensing periods~\cite{vtc/Xu20}. In practice, these models are often employed in a hybrid manner, in which CNNs are first used to extract spatial features from sensing data, followed by RNN or LSTM layers that capture temporal dynamics across consecutive sensing periods~\cite{icl/Xie00L20,access/Soni20}. Such temporal modeling capability allows the detector to incorporate historical sensing information when inferring current spectrum occupancy, which is particularly beneficial in dynamic environments with periodic or bursty PU transmissions. In CSS systems, LSTM-based models are commonly adopted to enhance local sensing reliability prior to information reporting~\cite{icl/Janu23}, or to assist fusion by aggregating temporally correlated sensing data from multiple sensors~\cite{tvt/Ding23}.

\subsubsection{Graph Neural Networks}
Graph neural networks (GNNs) are well suited for processing graph-structured data in non-Euclidean spaces and for capturing latent correlations among neighboring nodes. This property makes GNNs particularly attractive for CSS scenarios, in which SUs are geographically distributed and their sensing observations exhibit complex spatial dependencies.
A GNN-based spectrum sensing framework is proposed~\cite{tvt/Janu23}, where relationships among PU signals observed by multiple SUs are learned through a graph-based formulation. By operating on graphs with variable numbers of nodes, this approach naturally accommodates dynamic sensor participation. In this framework, energy detection is first performed to obtain received signal strength (RSS) measurements, and the adjacency matrix is then constructed based on these measurements. To enrich sensing representations, a multiview GNN is developed~\cite{iotj/Dong25} by integrating multiple manually designed features (e.g., RSS, phase-based statistics, and trajectory-related covariance features). While such handcrafted features can provide additional diversity, heavy reliance on manual design may limit adaptability in heterogeneous and time-varying sensing environments. To address this issue, a hybrid CNN-GNN architecture is introduced~\cite{tccn/Dong25}, where a CNN automatically extracts discriminative features from covariance matrices to form graph representations, thereby reducing dependence on handcrafted inputs. Beyond detection accuracy, the relational modeling capability of GNNs offers unique advantages in securing CSS. For instance, by leveraging GNN-based spatial interpolation to cross-verify sensing reports against the learned graph topology, trust evaluation mechanisms can effectively identify malicious nodes and mitigate data falsification attacks~\cite{twc/Zhang22}.

\subsubsection{Transformer Networks}
Distinct from the local receptive fields of CNNs, Transformer networks adopt a self-attention mechanism to model global correlations and long-range dependencies across high-dimensional sensing data. In single-node sensing, this architecture transcends the limitations of local feature extraction by treating temporal samples or spectrum bins as interactive tokens. This capability allows for the joint learning of temporal switching characteristics and inter-band spectrum correlations, achieving a favorable accuracy-complexity trade-off~\cite{icl/Chen25,twc/Zhang24}. Furthermore, by reformulating wideband sensing as a semantic segmentation task, the attention mechanism effectively disentangles multiple coexisting signals within complex spectrograms, significantly enhancing segmentation efficiency in multi-user environments~\cite{wcl/HuynhThe25,access/Kong25}. In the context of multiple sensors, Transformer networks fundamentally reshape cooperative fusion by facilitating cross-sensor interactions beyond simple aggregation. Through self-attention mechanisms, distributed sensing information is jointly modeled to capture spatial correlations among users~\cite{icl/Janu25,wcl/Fang25}. Such designs demonstrate superior robustness, particularly in mitigating the effects of imperfect reporting links and sensor mobility, thereby offering a scalable solution for dynamic CSS networks~\cite{wcl/Fang25}.

\subsubsection{Other Discriminative Models}
Beyond the aforementioned architectures, emerging discriminative models have been explored to address specific challenges in CSS, particularly focusing on computational latency, energy efficiency, and model compactness.
Addressing the computational complexity of long-sequence modeling, the Mamba (a representative state-space model) architecture offers a linear-complexity alternative to Transformer networks. By integrating CNN-based spatial feature extraction with Mamba-based selective state-space modeling, this approach facilitates parallelized processing, overcoming the sequential limitations of RNNs. This design significantly reduces training overhead while maintaining high accuracy in discerning dynamic spectrum patterns~\cite{mlsp/Yi24}.
For energy-constrained edge platforms, spiking neural networks (SNNs) provide a biologically inspired paradigm leveraging event-driven processing. Unlike continuous-value activations in standard DNNs, SNNs utilize sparse spikes to achieve substantial power savings. For instance, quantized SNNs on neuromorphic chips demonstrate high energy efficiency while preserving sensing accuracy~\cite{tgcn/Liu24}. Furthermore, their inherent noise robustness effectively distinguishes target signals from heavy co-channel interference with minimal power consumption~\cite{taes/Dakic24,wcnc/Dakic24}.
To further optimize the performance-complexity trade-off, Kolmogorov--Arnold networks (KANs) have emerged as a lightweight alternative to traditional DNNs. By employing adaptive activation functions on edges rather than nodes, KANs enable the construction of compact models with significantly fewer parameters~\cite{Pan_2024}. Moreover, such architectures exhibit superior interpretability and robustness, effectively suppressing noise to enhance detection reliability in low single-to-noise ratio (SNR)~\cite{comsnets/Patel25,jsen/Yan25}.

\subsection{Generative Deep Learning}
While discriminative models have achieved remarkable success in CSS, they fundamentally rely on large-scale labeled datasets to cover diverse signal characteristics and complex channel environments. In practice, acquiring comprehensive ground truth across heterogeneous spectral scenarios is often cost-prohibitive or infeasible. Distinct from discriminative approaches that directly approximate the decision boundary (i.e., the posterior $P(H_i|\mathbf{X})$) for classification, generative DL models focus on learning the underlying statistical distribution of spectrum observations (e.g., $P(\mathbf{X})$ or $P(\mathbf{X}|H_i)$). This paradigm shift offers two distinct advantages. First, by sampling from the learned distribution, these models can synthesize high-fidelity sensing data, serving as an effective mechanism for data augmentation in data-limited scenarios. Second, the learned generative priors can be directly leveraged for inverse problems, such as denoising and spectrum sensing.

\subsubsection{Variational Autoencoder}
Variational autoencoder (VAE) is probabilistic generative model that learn compact latent representations while imposing a structured distribution in the latent space. Typically trained without labels, VAE is attractive for spectrum sensing scenarios in which annotated samples are scarce or difficult to obtain. By mapping high-dimensional spectrum data into a low-dimensional probabilistic manifold, VAE-based approaches effectively capture the intrinsic statistical characteristics of PU signals, ensuring robust representation even under dynamic channel and interference conditions. Distinct from directly performing classification tasks, VAE-based methods typically reformulate sensing as a clustering task within the latent space. Representative studies adopt Gaussian mixture models~\cite{tvt/Xie20}, K-means clustering~\cite{tccn/Liao24}, or affinity propagation~\cite{globecom/Khalek24} to distinguish different PU states based on latent embeddings.

\subsubsection{Generative Adversarial Networks}
Generative adversarial networks (GANs) constitute a prominent class of generative models that approximate the underlying data distribution through a min-max game between a generator and a discriminator. The generator aims to synthesize realistic spectrum samples, while the discriminator seeks to distinguish generated samples from real observations, thereby driving the generator to capture statistical characteristics of PU signals.
In CSS, GANs are primarily employed to alleviate the scarcity of labeled training data, which is a long-standing challenge in practical spectrum sensing systems~\cite{icc/Davaslioglu18}. By learning the distribution of PU signals, GAN-based frameworks can generate high-fidelity synthetic samples for data augmentation, effectively enriching training datasets and improving the generalization performance of sensing models~\cite{wcl/Cai22,iwcmc/Liu21,tce/Gao24}. Such augmentation-based strategies have proven particularly beneficial in low-SNR regimes and heterogeneous signal environments, where the acquisition of labeled data is costly or infeasible~\cite{icl/Xu24,10841791}. 
Beyond data augmentation, GANs have been increasingly integrated into CSS pipelines to enhance sensing robustness under adverse conditions. In particular, GANs can act as powerful denoising and signal recovery modules, enabling the reconstruction of clean spectrum representations from noisy observations~\cite{jsen/Yan25}. Similar principles have been extended to compressive spectrum sensing, where GANs learn the low-dimensional manifold of spectrum signals and enable accurate reconstruction from undersampled measurements, thereby reducing sensing overhead while maintaining detection performance~\cite{icc/Meng20}. 
Furthermore, GAN-based approaches offer a potent solution for addressing distribution mismatch across sensing environments. By aligning feature distributions between source and target domains, GANs facilitate domain adaptation for CSS, improving robustness to variations in channel conditions, noise statistics, and signal types~\cite{icl/Zhao23}. Collectively, these studies highlight GANs as a versatile generative tool for enhancing data efficiency, robustness, and adaptability in CSS, complementing discriminative models in challenging and dynamic wireless environments.

\subsubsection{Diffusion Models}
Diffusion models have emerged as a powerful generative approach, capturing data distributions by progressively denoising a latent variable from a Gaussian prior. Unlike GANs and VAE, diffusion models avoid mode collapse and posterior approximation, leading to more stable training and better coverage of the data distribution~\cite{comsur/Fan26}. In the context of CSS, they are employed for spectrogram segmentation by generalizing conventional hypothesis testing into pixel-level inference. A representative framework, SpecDiff~\cite{wcl/HuynhThe26}, provides a concrete example by applying latent diffusion models to identify and disentangle coexisting new radio (NR), long-term evolution (LTE), and radar signals, effectively distinguishing them from background noise (i.e., vacant spectrum) within a shared environment. Rather than operating directly on raw inputs, it compresses the complex spectrogram into a compact latent space and employs a specialized noise prediction network. Specifically, this network integrates a frequency filter to decompose latent representations, effectively separating underlying stable signal structures from high-frequency noise. By leveraging attention-guided mechanisms, SpecDiff robustly recovers clean signal masks even under severe channel impairments, maintaining high identification accuracy at SNR as low as $-10$\,dB. Furthermore, it achieves state-of-the-art segmentation accuracy while remaining highly parameter-efficient compared to general-purpose models. Consequently, the ability to model spectrum distributions allows these methods to naturally yield robust occupancy decisions. This positions diffusion models as a promising generative alternative, particularly where fine-grained structural awareness and robustness in harsh environments are paramount.

\subsubsection{Generative Foundation Models}
Generative foundation models (GFMs) follow a unified ``pre-training plus fine-tuning'' paradigm and have recently attracted attention for wireless intelligence tasks. Most GFMs are built on Transformer-based architectures and are pre-trained with self-supervised objectives on large-scale datasets, allowing them to capture underlying signal distributions and output transferable representations~\cite{alikhani2025lwm}. Such representations can then be adapted to downstream tasks supporting a more general spectrum intelligence framework~\cite{javaid2025agi}. 
Recent efforts explore different ways of leveraging GFMs for spectrum sensing. The WirelessLLM framework~\cite{jcin/Shao24}, for instance, examines the direct use of off-the-shelf models through prompt engineering. Instead of updating model parameters, it relies on a few-shot prompting strategy, in which a small number of task examples are concatenated into the input prompt. Spectrum measurements are first tokenized, and the model performs inference via in-context learning. This allows the same model to handle multiple tasks, such as sensing and resource allocation, while maintaining competitive detection performance compared with classical ED method. 
A different direction is taken by SpectrumFM~\cite{jsca/Zhou25}, which builds a domain-specific foundation model tailored to spectrum data. Its training starts from a large curated dataset covering diverse modulation formats under varying SNR conditions. The pre-training stage combines masked reconstruction with next-slot prediction, encouraging the model to learn both local signal patterns and temporal dependencies. For downstream adaptation, parameter-efficient fine-tuning (PEFT) is adopted to reduce computational overhead. With this design, SpectrumFM reports an AUC of $0.97$ at $-4$\,dB SNR, outperforming CNN- and LSTM-based baselines. In general, GFMs represent a shift away from strictly task-specific designs toward more flexible and reusable models, offering a promising basis for spectrum sensing in heterogeneous environments.

\begin{table*}[!t]
    \centering
    \caption{Performance Comparison of Discriminative DL models for CSS}
    \label{tab:discriminative}
    \begingroup
    \belowrulesep=0pt
    \aboverulesep=0pt
    \footnotesize
    \tabcolsep 4pt
    \renewcommand{\arraystretch}{1.4}
    \begin{tabularx}{\textwidth}
    {@{\hspace*{3pt}}
    >{\hsize=0.4\hsize\bfseries\arraybackslash}X 
    >{\hsize=0.8\hsize\raggedright\arraybackslash}X 
    >{\hsize=1.1\hsize\raggedright\arraybackslash}X 
    >{\hsize=0.4\hsize\centering\arraybackslash}X 
    >{\hsize=1.3\hsize\raggedright\arraybackslash}X 
    @{\hspace*{3pt}}}
        \toprule
        Model & \textbf{Primary Mechanism} & \textbf{Pros and Cons} & \textbf{Complexity} & \textbf{Typical Scenarios} \\
        \midrule
        CNNs & Spatial feature extraction & Strong 2D pattern recognition; limited long-range dependency modeling & Moderate & Sensing tasks utilizing signal covariance matrices and wideband spectrograms \\
        RNNs & Sequential dependency modeling & Effective temporal dynamics capture; high latency for long sequences & Moderate & Continuous tracking of PU temporal activities\\
        GNNs & Relational graph modeling & Adaptability to irregular topologies; high computational overhead for dynamic updates & High & Distributed sensing in sensor networks with irregular topologies \\
        Transformer & Global self-attention & Global contextual awareness; quadratic computational complexity & Very High & Complex environments with PU temporal activities and inter-band spectrum correlations \\
        Mamba & Selective state-space modeling & Linear scalability for long sequences; lower performance relative to Transformer networks & High & Low-latency and long-term PU activity monitoring \\
        SNNs & Sparse event-driven spikes & Ultra-low power consumption; hardware dependency and limited accuracy & Low & Ultra-low power edge sensing on neuromorphic devices \\
        KANs & Spline-based activations & High parameter efficiency; high sensitivity to spline grid density & Low & Resource-constrained sensing on edge IoT devices \\
        \bottomrule
    \end{tabularx}
    \endgroup
\end{table*}

\subsection{Deep Reinforcement Learning}
Beyond one-shot inference that learns $\mathcal{F}(\cdot;\boldsymbol{\alpha})$ and $\mathcal{G}(\cdot;\boldsymbol{\beta})$ and applies them independently in each sensing period, DRL adopts a sequential decision-making paradigm that optimizes the sensing operation and cooperation configuration across successive sensing periods, thereby shaping the observations and reports that serve as inputs to $\mathcal{F}(\cdot;\boldsymbol{\alpha})$ and $\mathcal{G}(\cdot;\boldsymbol{\beta})$. Each period $t$ comprises a sensing slot that produces local observations $\mathbf{X}_k^{t}$ and a reporting slot that yields the global decision $\hat{H}^{t}$. Since historical outcomes and resource consumption can affect subsequent configurations and thus the resulting observations, CSS naturally lends itself to a Markov decision process formulation and is often modeled as partially observable, where a policy is learned to act based on the information available up to period $t-1$.

A DRL formulation specifies a state, an action, and a reward that captures long-term objectives. The state summarizes the sensing history and system status, including recent occupancy decisions $\hat{H}^{t-1}$, observation statistics derived from $\mathbf{X}_k^{t-1}$, link-quality indicators in Eq.\,\eqref{eq:report}, and resource budgets such as energy or latency. The action configures the CSS pipeline before or during period $t$, for example by selecting participating sensors, allocating sensing and reporting resources, choosing the sample size $N$, and selecting channels to sense. DRL adapts configuration variables that govern which observations are collected and how reports are formed, thus shaping the inputs to $\mathcal{F}(\cdot;\boldsymbol{\alpha})$ and $\mathcal{G}(\cdot;\boldsymbol{\beta})$ over time~\cite{icoin/Pham23}. 
Depending on whether decision making is centralized at the FC or distributed across sensors, existing DRL-based CSS studies can be broadly categorized into single-agent DRL and multi-agent DRL.

\subsubsection{Single-Agent Deep Reinforcement Learning}
In the single-agent DRL framework, a centralized controller acts as the global decision-maker, optimizing sensing and reporting configurations across successive periods. Distinct from static scheduling protocols, this framework models CSS as a sequential decision process, in which the agent dynamically refines its belief state to balance global detection accuracy against overall resource consumption. 
A representative application of this paradigm addresses the bottleneck of reporting overhead~\cite{icl/Sarikhani20}. Instead of activating all sensors, the FC employs a deep Q-network (DQN) to sequentially activate sensors based on their potential information gain, stopping the query process once a reliable decision can be made. By incorporating spatial correlation derived from sensor location information into the reward structure, the learned policy prioritizes sensors with unique perspectives, effectively filtering out redundant data streams. Consequently, this approach maintains detection performance comparable to full-cooperation schemes while drastically reducing the number of active sensors. Further advances~\cite{ieicetb/Xu25} extend this logic by adopting dueling double DQN architectures to achieve a superior balance between detection reliability and collaboration costs.

\subsubsection{Multi-Agent Deep Reinforcement Learning}
In the multi-agent DRL framework, CSS is formulated as a distributed control problem, where each sensor operates as an independent agent, making local sensing and cooperation decisions based on partial observations and limited message exchange. Compared with centralized control, multi-agent DRL facilitates scalable decision-making and decentralized execution, enabling coordination to emerge through shared rewards and structured interactions.
To reduce sensing energy consumption, a hierarchical multi-agent DRL scheme, termed PDSRNN, is proposed~\cite{wcl/Ngo23}. In this scheme, sensor agents perform sensing only during the first slot of a frame and utilize an RNN integrated with post-decision state learning to predict occupancy for the subsequent frame. Their local decisions are then aggregated by a central FC agent to form a reliable global spectrum decision, enabling efficient cooperative sensing.
Addressing the multi-channel selection problem, some studies~\cite{icl/Gao21,tvt/Gao24} employ multi-agent deep deterministic policy gradient (MADDPG) to guide sensors in collaboratively identifying and accessing idle channels. A further advancement~\cite{tvt/Gao24} integrates an attention mechanism into MADDPG, allowing each sensor to dynamically weigh information from candidate partners. By learning partner importance based on historical reliability and geographical correlation, this mechanism enhances both sensing accuracy and system interpretability under dynamic network conditions.
Building on the concept of reliability-based collaboration, a subsequent study~\cite{vtc/Wang25} integrates a trust-aware weighted fusion mechanism with the multi-agent proximal policy optimization (MAPPO) algorithm. The fusion mechanism dynamically adjusts each sensor's contribution to the global decision based on historical accuracy, while MAPPO enables effective centralized training for distributed execution. Together, these components significantly improve detection performance in dynamic environments.

In essence, DRL complements representation learning by extending optimization from instantaneous inference to sequential control. It provides an adaptive layer that dynamically orchestrates sensing, reporting, and collaboration strategies, ensuring robust performance under temporal dynamics and resource constraints.

\begin{figure*}[!t]
    \centering
    \includegraphics[width=0.8\linewidth]{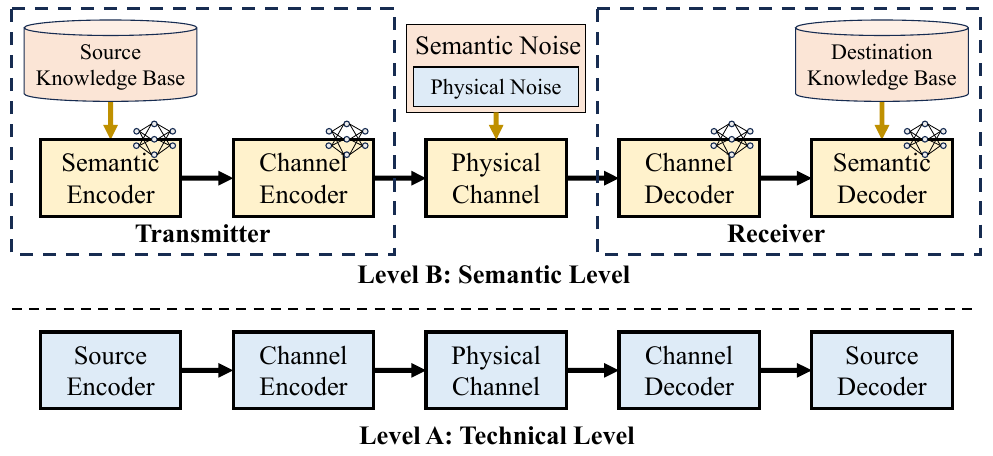}
    \caption{Comparison of architectures between conventional communication and SemCom.}
    \label{fig:semcom}
\end{figure*}

\subsection{Lessons Learned}
In this section, we present an overview of AI-driven CSS methodologies across three learning paradigms. The lessons learned are as follows:
\begin{itemize}
    \item Discriminative DL has shifted CSS from manual feature engineering toward automated pattern recognition. Tab.\,\ref{tab:discriminative} provides a performance comparison of these discriminative models in terms of their primary mechanisms, pros and cons, complexity, and preferred applicability. The progression from CNNs and RNNs to GNNs and Transformers demonstrates a growing ability to model complex spatio-temporal relationships. On the other hand, lightweight models (e.g., SNNs and KANs) address the computational and energy constraints inherent in edge computing. 
    \item Generative DL excels at capturing the data distribution of wireless signals. VAE and GANs are commonly used to assist discriminative models in enhancing detection, whereas diffusion models are applied independently for spectrogram segmentation. Furthermore, although GFMs support a broad spectrum of wireless intelligent tasks, deploying them for CSS often incurs high computational costs and fails to outperform well-designed task-specific discriminative models.
    \item DRL addresses CSS by optimizing long-term performance through sequential interactions with the environment. The sensing policy is learned to maximize cumulative rewards, leading to adaptive strategies for sensor selection, channel access, and energy management. Extensions to multi-agent DRL enable distributed implementations, but also introduce additional challenges in achieving stable convergence under dynamic conditions.
\end{itemize}

\section{Semantic Communication-Enabled Cooperative Spectrum Sensing}
\label{Sec.SemCom}
In this section, we introduce an AI-native communication paradigm, namely SemCom. We then apply SemCom to CSS, revisit the problem from the SemCom perspective, and discuss scenarios involving both single-user and multi-user setups.

\begin{figure}
    \centering
    \includegraphics[width=0.9\linewidth]{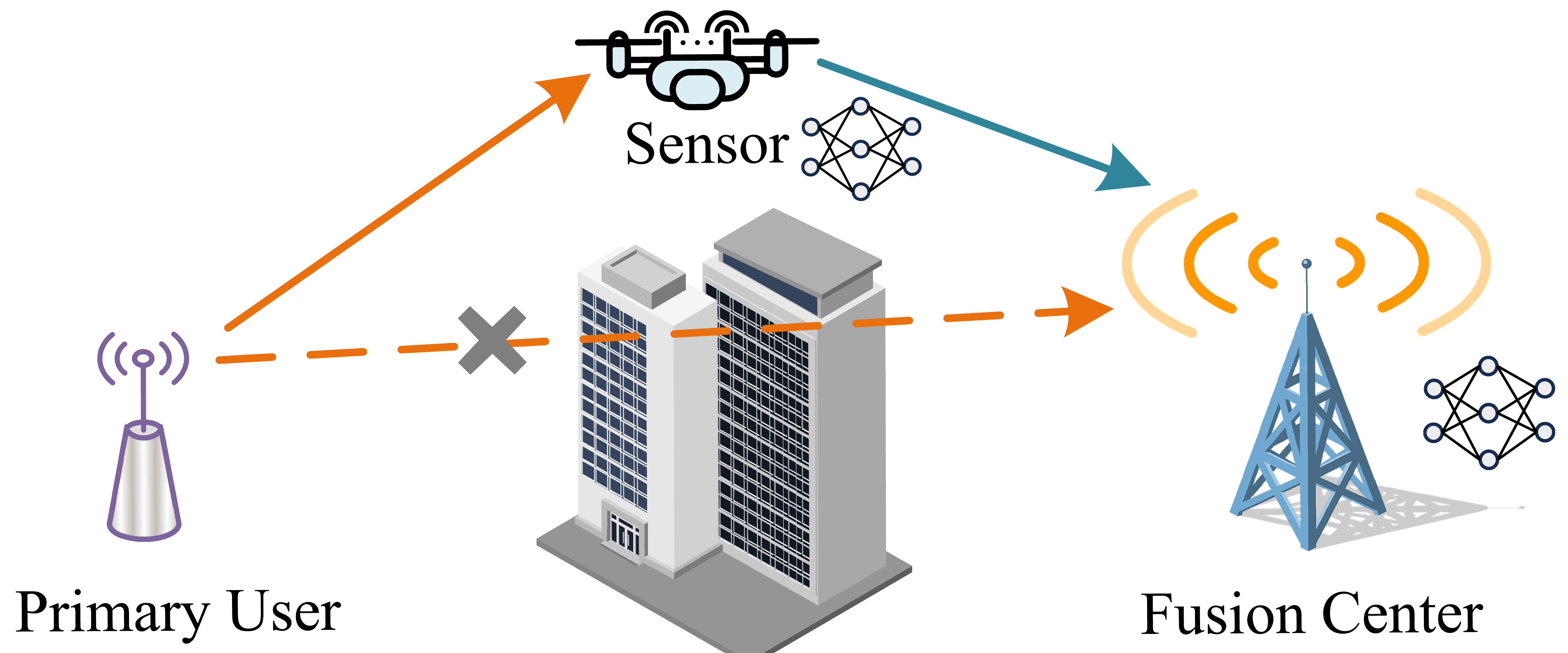}
    \caption{System model of single-user SemCom for remote spectrum sensing. The sensor encodes local observations into semantic representations via a semantic encoder and transmits them to the FC, where a semantic decoder infers the PU status.}
    \label{fig:SU-SemCom}
\end{figure}

\begin{figure*}[!t]
    \centering
    \subfloat[Multi-user SemCom without AirComp (OMA-based reporting).]{
        \includegraphics[width=0.6\linewidth]{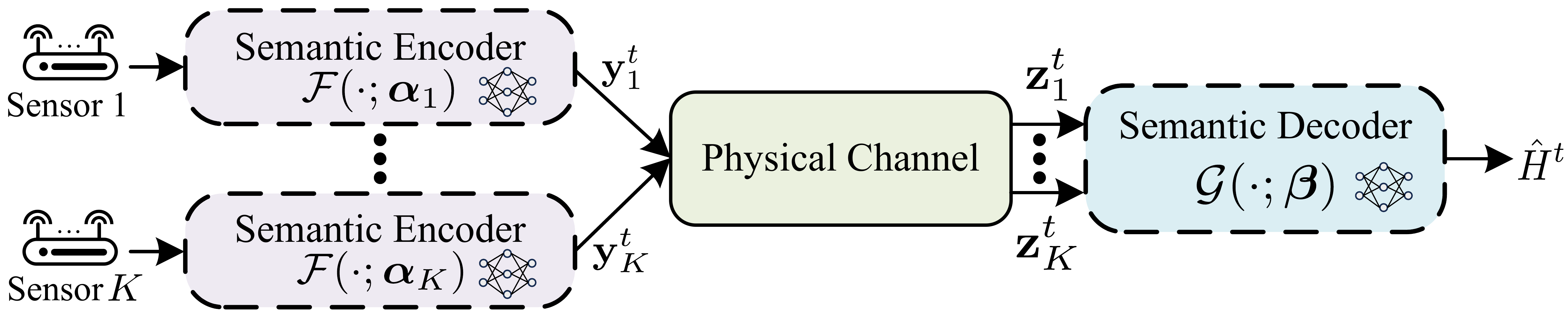}
        \label{fig:MU-SemCom-OMA}
    }
    \\
    \subfloat[Multi-user SemCom with AirComp.]{
        \includegraphics[width=0.6\linewidth]{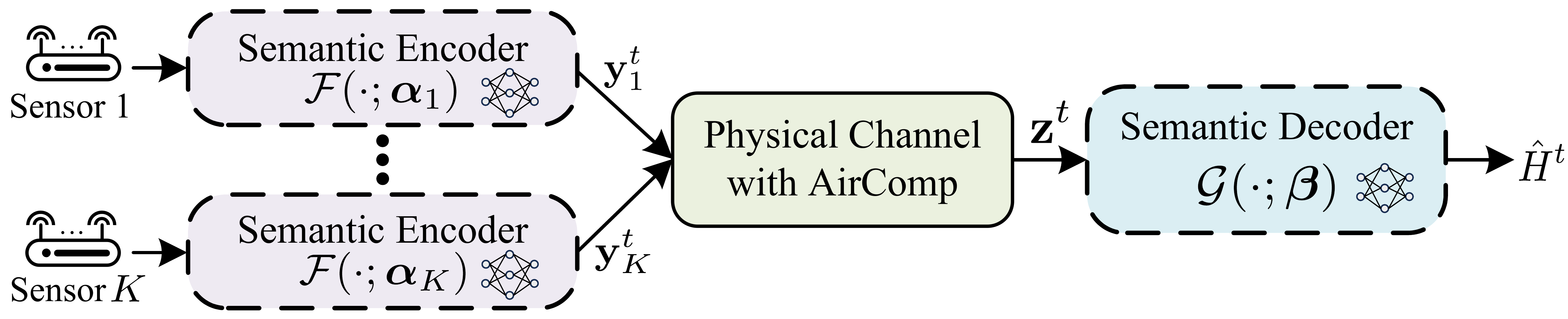}
        \label{fig:MU-SemCom-AirComp}
    }
    \caption{System models of multi-user SemCom. (a) Traditional OMA-based reporting, utilizing orthogonal resources (e.g., time or frequency), which results in communication overhead (latency or bandwidth) scaling linearly with the number of sensors. (b) AirComp-based reporting, where semantic representations are concurrently aggregated over the air on the same resource blocks, achieving high spectral efficiency and scalable fusion independent of the network size.}
    \label{fig:MU-SemCom}
\end{figure*}

\subsection{Concept and Motivation}
Conventional wireless communication systems are fundamentally rooted in the Shannon paradigm, in which the primary objective is the reliable reproduction of transmitted bits. In this bit-centric paradigm, communication is treated as a content-agnostic transmission layer, largely decoupled from upper-layer applications and inference tasks. While highly effective for general data delivery, this philosophy inevitably leads to substantial redundancy in task-oriented scenarios, as massive amounts of raw data are transmitted regardless of their semantic relevance or utility for downstream tasks~\cite{jcin/Lan21}. 
To transcend this limitation, SemCom introduces a paradigm shift from bit reproduction to meaning delivery. The architectural contrast between conventional communication and SemCom is summarized in Fig.\,\ref{fig:semcom}. Revisiting Weaver's classical classification~\cite{etc/Weaver53}, communication problems are conceptualized across three levels: technical, semantic, and effectiveness. While conventional systems primarily address the technical level (Level A), SemCom targets the semantic (Level B) and effectiveness (Level C) levels, aiming to ensure that the receiver precisely interprets the intended meaning or successfully accomplishes the target tasks~\cite{comsur/Lu24}. Under this perspective, communication performance is measured by semantic correctness or task performance, instead of purely bit-level fidelity. 

Enabled by recent breakthroughs in DL, SemCom systems have become practically realizable~\cite{tsp/Xie21}. By adopting end-to-end learning paradigms and leveraging the powerful representation capabilities of DNNs, SemCom integrates feature extraction, representation learning, and resilient transmission into a unified, task-oriented framework~\cite{comsur/Yang23,comsur/Nguyen26}. Specifically, rather than transmitting raw observations, the transmitter utilizes a DNN to distill input data into compact semantic representations, preserving task-critical features while eliminating redundancy. Correspondingly, the receiver employs a paired decoder to recover the intended meaning from the corrupted signals, which take the form of semantic-level source reconstruction or task-specific decisions. Furthermore, unlike the ``cliff effect'' characteristic of separated source-channel coding schemes, these jointly optimized representations exhibit inherent robustness, ensuring graceful degradation across diverse SNR regimes. As a result, SemCom emerges as an AI-native communication paradigm that tightly couples transmission with inference objectives.

n the context of CSS, the reporting channel constitutes a critical bottleneck, imposing a fundamental conflict between information fidelity and transmission efficiency. Conventional approaches typically address this trade-off through two directions: transmitting high-dimensional raw observations $\mathbf{X}^t_k$, or reporting low-bit quantized decisions. While the former preserves rich sensing features, it incurs prohibitive bandwidth overhead and suffers severe distortion under noisy reporting conditions. Conversely, the latter achieves bandwidth efficiency via aggressive quantization, yet inevitably discards decision-relevant information, which degrades the overall detection performance. 
Therefore, CSS serves as a representative scenario for SemCom, which adopts a task-oriented paradigm by encoding observations $\mathbf{X}^t_k$ into compact, noise-resilient semantic representations $\mathbf{y}^t_k$. From the information bottleneck perspective, the objective is to maximize $I(\mathbf{y}_k; H)$ while minimizing $I(\mathbf{y}_k; \mathbf{X}_k)$. 
Therefore, CSS serves as a representative scenario for SemCom, which adopts a task-oriented paradigm by encoding observations $\mathbf{X}^t_k$ into compact, noise-resilient semantic representations $\mathbf{y}^t_k$. In such a case, semantic information refers to the task-relevant features of the observed signals that are sufficient for accurately inferring the PU status, while discarding redundant and task-irrelevant details. From the information bottleneck perspective, the objective is to maximize $I(\mathbf{y}_k; H)$ while minimizing $I(\mathbf{y}_k; \mathbf{X}_k)$.

\subsection{Single-User Semantic Communication}
Single-user SemCom, illustrated in Fig.\,\ref{fig:SU-SemCom}, instantiates the general framework in Section\,\ref{SubSec.GeneralFramework} as a task-oriented point-to-point link. In this context, the local processing function $\mathcal{F}(\cdot;\boldsymbol{\alpha})$ in Eq.\,\eqref{eq:encoding} serves as a semantic encoder, mapping the raw observations into a compact representation $\mathbf{y}^{t}$ that is resilient to channel impairments. Correspondingly, the fusion function $\mathcal{G}(\cdot;\boldsymbol{\beta})$ in Eq.\,\eqref{eq:decoding} acts as a semantic decoder at the FC to directly infer the spectrum occupancy $\hat{H}^{t}$ from the received noisy signal $\mathbf{z}^{t}$. Unlike conventional separate source-channel coding, $\mathcal{F}(\cdot;\boldsymbol{\alpha})$ and $\mathcal{G}(\cdot;\boldsymbol{\beta})$ are jointly optimized to maximize detection accuracy under the constraints of the reporting channel. 
Specifically, this can be formulated as minimizing a task-specific loss function, such as cross-entropy $\mathcal{L}_{CE}(\cdot, \cdot)$ between true label $H^t$ and $\hat{H}^t$:
\begin{equation}
    \min_{\boldsymbol{\alpha}, \boldsymbol{\beta}} \mathbb{E} \left[ \mathcal{L}_{CE} \left( H^t, \mathcal{G}\left( \mathcal{C}\left( \mathcal{F}\left(\mathbf{X}^t;\boldsymbol{\alpha}\right) \right); \boldsymbol{\beta} \right) \right) 
    \right],
\end{equation}
where the expectation is taken over sensing noise, PU signal statistics, and reporting channel variations.
A representative work~\cite{mlsp/Yi22} implements this architecture using a deep autoencoder for remote spectrum sensing. By treating the reporting channel as a non-trainable layer within the neural network, the encoder learns to extract and protect decision-critical features (e.g., low-dimensional manifolds of covariance matrices) against noise and fading. Experimental evidence suggests that such end-to-end learning effectively eliminates the ``cliff effect'' observed in digital transmission, maintaining robust detection performance even in low SNR regimes where quantization-based methods fail.

\begin{figure*}[t]
    \centering
    \subfloat[$P_d$ versus $\widetilde{SNR}$ in sensing channels when $\widehat{SNR}=-3$\,\text{dB} \label{fig:a}]{
        \includegraphics[width=0.4\textwidth]{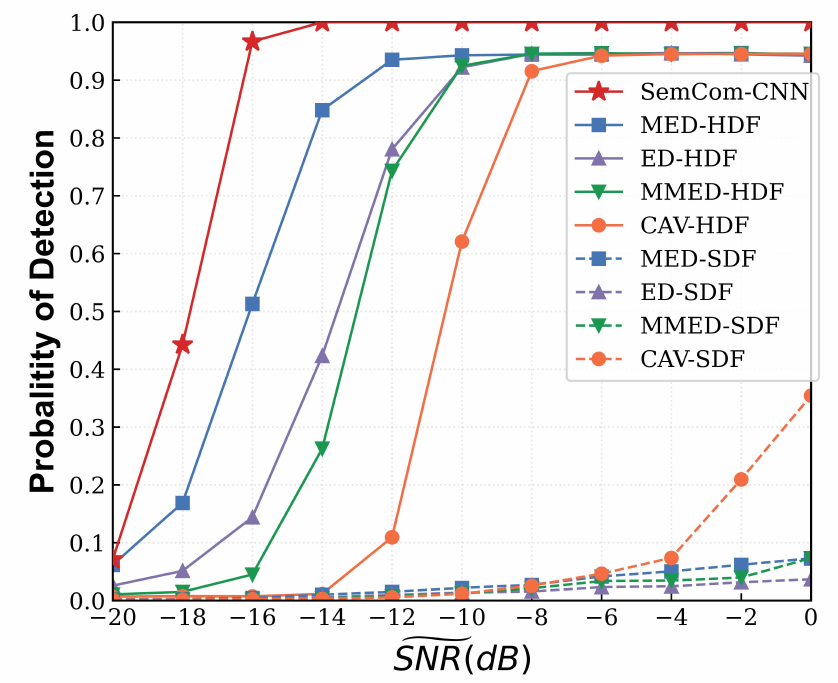}
    }
    \hspace{12pt}
    \subfloat[$P_d$ versus $\widehat{SNR}$ in reporting channels when $\widetilde{SNR}=-15$\,\text{dB}  \label{fig:b}]{
        \includegraphics[width=0.4\textwidth]{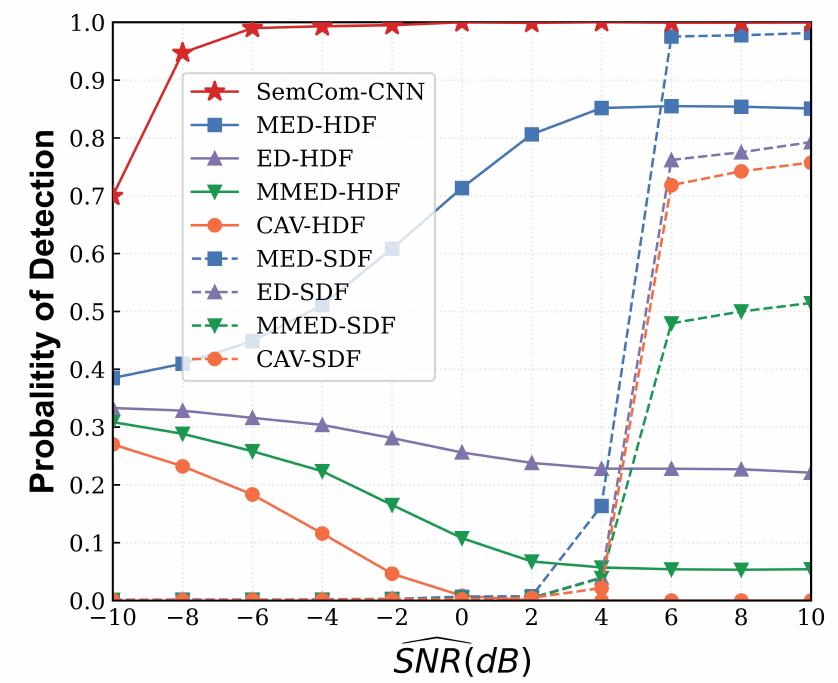}
    }
    \caption{Performance comparison.}
    \label{fig:performance}
\end{figure*}

\begin{table*}[!t]
    \belowrulesep=0pt
    \aboverulesep=0pt
    \renewcommand{\arraystretch}{1.4}
    \doublerulesep 2.2pt
    \centering
    \caption{Comparison in one sensing period, when $K=6$, $M=28$, $N=100$, $\widetilde{SNR}=-15\,\text{dB}$, $\widehat{SNR}=0\,\text{dB}$.}
    \label{Tab.methods}
    {\footnotesize
    \begin{tabular*}{\textwidth}{@{\extracolsep{\fill}}c*9{c}}
    \toprule
     & \multicolumn{4}{c}{HDF (Majority Rule)} & \multicolumn{4}{c}{SDF (Equal Gain Combining)} & \multirow{2}{*}{SemCom-CNN} \\
    \cmidrule(lr){2-5}\cmidrule(lr){6-9}
    Metrics / Methods & ED & MED & MMED & CAV & ED & MED & MMED & CAV &  \\
    \midrule
    $P_d$ ($P_{fa}=10^{-3}$) & 0.260 & 0.698 & 0.129 & 0.042 & 0.002 & 0.005 & 0.002 & 0.003 & 0.995 \\
    \midrule
    Number of utilized subchannels & 6 ($K$) & 6 ($K$) & 6 ($K$) & 6 ($K$) & 48 ($8\times K$) & 48 ($8\times K$) & 48 ($8\times K$) & 48 ($8\times K$) & 8 \\
    \midrule
    Inference Time (ms) & 0.091 & 1.658 & 1.659 & 0.216 & 0.091 & 1.658 & 1.659 & 0.216 & 0.794 \\
    \bottomrule
    \end{tabular*}}
\end{table*}

\subsection{Multi-User Semantic Communication}
Multi-user SemCom, as depicted in Fig.\,\ref{fig:MU-SemCom}, extends the single-user SemCom paradigm to distributed networks, where multiple sensors collaboratively contribute semantic representations for a shared inference task~\cite{jsac/Xie22}. Rather than reporting raw signal samples, each sensor encodes its local observations into a task-oriented semantic vector $\mathbf{y}_k^{t}$ that preserves decision-relevant information while reducing the reporting overhead. The FC then performs joint inference by fusing the received semantic messages. Such semantic-level collaboration exploits spatial diversity and can improve robustness to sensing uncertainty and reporting impairments.
In the conventional orthogonal multiple access (OMA)-based realization in Fig.\,\ref{fig:MU-SemCom-OMA}, sensors occupy orthogonal time- or frequency-domain resources to avoid mutual interference. Consequently, the reporting overhead scales linearly with $K$: enforcing orthogonality across time increases the reporting latency as $K$ grows, whereas enforcing orthogonality across frequency increases the required bandwidth proportionally to $K$.

To break this scaling, over-the-air computation (AirComp) can be integrated into multi-user SemCom, as shown in Fig.\,\ref{fig:MU-SemCom-AirComp}~\cite{wc/Luo24}. By leveraging the superposition property of the wireless channel, AirComp enables concurrent transmissions and aggregates semantic representations directly over the air. Consequently, the received signal at the FC becomes a weighted sum of the individual semantic representations, formulated as:
\begin{equation}
    \label{eq:report_ota}
    \mathbf{z}^{t} = \sum_{k=1}^K \widehat{h}_k^{t} {b}_k^{t} \mathbf{y}_k^{t} + \widehat{\mathbf{v}}^{t},
\end{equation} 
where ${b}_k^{t}$ is a precoding scalar to overcome the negative impact of fading. 
This analog aggregation reduces the required reporting resources (time or bandwidth) for the aggregation process, making the overall reporting overhead largely independent of the network size $K$.
Tailoring this to CSS, an integrated communication and computation framework is proposed~\cite{tcom/Yi25}. It jointly optimizes distributed SemCom across sensors and the FC to minimize the transmitted data volume without degrading detection performance, and exploits AirComp to further reduce spectrum usage in the reporting phase via concurrent aggregation. Under i.i.d. PU signal samples, the resulting design is theoretically shown to be equivalent to the optimal E-C detector with equal-gain SDF, and numerical results demonstrate improved robustness to SNR variations in both sensing and reporting channels while scaling favorably with the sample size and the number of sensors.
Building on this line of work, temporal modeling based on the Mamba architecture has been incorporated~\cite{mlsp/Yi24}, demonstrating that exploiting both spatial synergy and temporal correlations of PU activity can significantly enhance sensing reliability in dynamic environments.

To demonstrate the potential of the SemCom-enabled CSS framework, a lightweight CNN-based SemCom model, referred to as SemCom-CNN, is developed. 
We consider a scenario with $K=6$ sensors, each equipped with $M=28$ antennas and $N=100$ baseband samples per sensing period. The sensing and reporting channel SNRs are denoted by $\widetilde{SNR}$ and $\widehat{SNR}$, respectively. The reporting channel is modeled as a Rician channel with a $K$-factor of $0$\,dB. 
Fig.\,\ref{fig:performance} presents the detection probability $P_d$ versus $\widetilde{SNR}$ and $\widehat{SNR}$ for different algorithms. The SDF-based methods operate reliably only when $\widehat{SNR} \geq 6$\,dB, as channel impairments introduce severe reporting errors that significantly degrade its performance. In contrast, under very low $\widehat{SNR}$, HDF-based methods suffer from nearly random decisions at the FC, leading to a performance floor around $0.34$~\cite{tcom/Yi25}. As $\widehat{SNR}$ increases, all methods gradually approach their performance under ideal reporting channels. However, the relatively poor performance of CAV, MMED, and ED persists due to their intrinsic limitations at low sensing SNR (e.g., $\widetilde{SNR} = -15$\,dB). 
In particular, the proposed SemCom-CNN method demonstrates strong robustness against both $\widetilde{SNR}$ and $\widehat{SNR}$ variations.
Tab.\,\ref{tab:comparison} summarizes the communication and computational overhead of different methods. SemCom achieves low communication overhead and, when combined with AirComp, requires a constant number of subchannels regardless of the number of sensors. Moreover, its inference latency remains comparable to that of conventional methods.

\subsection{Lessons Learned}
SemCom-enabled CSS represents a joint communication and computation framework that shifts from bit-centric reporting to task-oriented intelligence sharing. The lessons learned are as follows:
\begin{itemize}
    \item By jointly optimizing the encoder and decoder, SemCom significantly reduces the transmitted data volume and avoids the ``cliff effect,'' maintaining robust detection performance even under severe channel impairments.
    \item SemCom is not separate from the AI-driven methodologies in Section\,\ref{Sec.AI}; rather, it provides a task-oriented communication framework that leverages models such as CNNs, Transformers, and Mambas for semantic encoding and decoding, enabling their representation learning capabilities to operate under wireless channel constraints.
    \item Multi-user SemCom with AirComp enables the concurrent aggregation of semantic features directly over the air. This transforms the reporting channel from a bottleneck into a functional computation layer, achieving $O(1)$ communication overhead with respect to the number of sensors and making large-scale collaborative sensing feasible.
\end{itemize}

\section{Limitations, Challenges and Open Research Problems}
\label{Sec.Challenges}
While AI-driven techniques and SemCom have demonstrated remarkable potential in enhancing wireless communication systems, their practical deployment in complex and dynamic environments remains challenging. This is particularly evident in CSS, in which AI models are required to operate reliably under heterogeneous devices, time-varying channels, and adversarial conditions. In this section, we discuss several key limitations and open research problems.

\subsection{Generalizability and Robustness} 
Most existing AI models for wireless applications are designed for specific tasks and operating conditions, which significantly limits their generalizability to unseen scenarios. For instance, a DNN trained for CSS is typically not transferable to other signal intelligence tasks such as modulation classification or specific emitter identification without substantial retraining. Moreover, AI-based detectors trained on simplified or simulated channel models suffer performance degradation when deployed in real-world CSS scenarios involving unknown interference, hardware impairments, or non-stationary environments.

A few works have explored this issue and proposed corresponding solutions. A general framework~\cite{alikhani2025lwm} is developed and pre-trained in a self-supervised manner on large-scale wireless channel datasets, providing representations for a wide range of downstream tasks in wireless communication and sensing systems. This study demonstrates that spatial and propagation characteristics learned from Sub-6 GHz channels can generalize to millimeter wave (mmWave) beam prediction. Another work~\cite{jsca/Zhou25} leverages diverse signal datasets for pre-training to support multiple downstream tasks, including automatic modulation classification, wireless technology classification, CSS, and anomaly detection. 
However, both approaches~\cite{alikhani2025lwm,jsca/Zhou25} introduce additional inference overhead, and there remains room for further performance improvement. Moreover, the continuous nature of wireless signals motivates the use of diffusion-based GFMs to better capture signal dynamics, thereby enabling more resilient and transferable spectrum intelligence.

\subsection{Explainability and Trustworthiness}
Despite the superior empirical performance of AI models, their black-box nature poses significant challenges for their deployment in safety-critical wireless applications~\cite{access/Chamola23}. This lack of transparency not only obscures the underlying decision-making logic but also masks potential biases and vulnerabilities, thereby undermining the fundamental pillars of trustworthy AI: reliability, fairness, and assurance~\cite{access/Chamola23}. In CSS scenarios, for instance, blind reliance on uninterpretable DNNs is risky, because network operators cannot discern whether a spectrum access decision stems from valid signal features or misleading correlations, leading to potential harmful interference to PUs. Consequently, bridging the gap between data-driven efficiency and human-understandable logic is imperative. 
Future research should focus on physics-informed learning paradigms that explicitly embed electromagnetic wave propagation models (e.g., large-scale path loss and multipath fading statistical priors) as regularization terms within the DNN loss functions. In terms of post-hoc explainability, new methods are needed to map the opaque attention weights of DL models back to human-interpretable spectrum features, such as cyclostationary signatures or energy statistics.

\subsection{Efficiency and Sustainability}
While advanced AI models can significantly improve task performance, their deployment often incurs substantial computational and energy costs, which conflict with the sustainability objectives of future 6G networks. In CSS scenarios, resource-constrained sensing nodes may be required to run complex DNNs, leading to excessive energy consumption and processing latency. This challenge reflects a fundamental trade-off between task performance and resource efficiency in AI-native wireless systems. As a result, sustainable AI paradigms that adapt model complexity to task requirements are of growing importance~\cite{pieee/Mao24}. 
An important direction for future research is the development of communication-aware efficiency techniques. For instance, mixture-of-experts (MoE) architectures can be designed to selectively activate experts according to channel state information (CSI), dynamically employing complex experts under severe fading conditions while relying on lightweight ones in high-SNR regimes. 
Furthermore, in SemCom-enabled CSS, semantic importance-aware structured pruning can be explored to retain only the neural pathways that are most critical for determining the PU's status. These strategies can enable energy-efficient intelligence at the wireless edge without sacrificing performance.

\subsection{Security and Privacy}
The integration of AI and SemCom into CSS also introduces new security and privacy vulnerabilities. AI models are susceptible to adversarial attacks, data poisoning, and model manipulation, which can severely undermine their reliability~\cite{cm/Wasilewska23}. For example, malicious users may intentionally transmit falsified sensing information, leading to incorrect cooperative decisions and spectrum misuse. One potential solution is to leverage blockchain to develop trust evaluation algorithms~\cite{iotj/Wang25}. In addition, training AI models for CSS often relies on data collected from distributed devices, which may implicitly reveal sensitive information such as user locations, mobility patterns, or activity states. This raises critical concerns regarding data privacy and secure collaboration. Promising directions include federated learning and privacy-aware semantic encoding. Federated learning enables distributed model training without sharing raw sensing data, where only model updates are exchanged and can be further protected via secure aggregation or differential privacy. In parallel, privacy-aware semantic encoding can be designed to selectively preserve task-relevant information while suppressing sensitive attributes (e.g., location or identity) through feature disentanglement.

\section{Conclusion}
\label{Sec.Conclusion}
In this paper, we have presented a comprehensive overview of CSS, reviewing its fundamental principles and recent advances enabled by AI. By using CSS as a representative case study, we have illustrated how AI-driven techniques are reshaping spectrum sensing paradigms and, more broadly, wireless communication systems. In particular, we have highlighted SemCom as a promising framework for enabling efficient and intelligent CSS, by shifting the focus from raw data exchange to task-oriented semantic information sharing. Looking ahead, with the rapid evolution of AI models, CSS serves as a compelling example of how AI can be effectively integrated into wireless communications. We hope that this work provides useful insights and guidance for future research on AI-driven wireless systems, and inspires further exploration of SemCom for next-generation spectrum sensing and beyond.


\bibliographystyle{IEEEtran}
\bibliography{References.bib}

\end{document}